# Direct current plasma spraying of mechanofused alumina-steel particles

M. Bouneder, H. Ageorges, M. El Ganaoui, B. Pateyron, P. Fauchais,

University of Limoges, SPCTS, 123 Avenue Albert Thomas, 87060 Limoges Cedex, France
*Tel. 33(0)555457505 / Fax. 33(0)555457211 / E-mail :* helene.ageorges@unilim.fr

**Abstract** :

Stainless steel particles (60 µm in mean diameter) cladded with an alumina shell (2 µm thick and manufactured by mechanofusion) were sprayed with an Ar-$H_2$ (53-7 slm) d.c. plasma jet (I = 500 A, P = 28 kW, $\rho_{th}$ = 56 %). Two main types of particles were collected in flight, as close as 50 mm downstream of the nozzle exit: particles with a steel core with pieces of alumina unevenly distributed at their surface and those consisting of a spherical stainless steel particle with an alumina cap. To understand the phenomena, stainless steel particles were also sprayed in the same conditions. Particles collected contained oxide nodules inside them (due to the convective movement induced by the plasma flow within the liquid particles) and presented also an oxide cap and an oxide shell (only observed between 70 and 100 mm downstream of the nozzle exit).
The plasma flow was modeled by a 2D steady parabolic model and a single particle trajectory by using the 3D Boussinesq-Oseen-Basset equation. The heat transfer, within the two-layer, stainless steel cladded by alumina, particle, considered the heat propagation phenomena including phase changes. The models allowed determining the positions, along the particle trajectory, where the convective movement could occur as well as the entrainment of the liquid oxide (the spinel for the pure stainless steel particles and the alumina for the cladded ones) to the leading edge of the in-flight particles. The heat transfer calculations showed the importance of the thermal contact resistance TCR between alumina and steel, which values were varied between $10^{-6}$ and $10^{-8}$ $m^2 \cdot K \cdot W^{-1}$ because no experimental value was available. However whatever maybe the TCR value, the alumina shell melting occurs always at the position, along the particle trajectory, where the force applied by the plasma flow onto the liquid oxide light layer is sufficient for its entrainment towards the leading edge of the particle (cap formation).

## 1. Introduction
In atmospheric plasma spraying, many parameters can influence the coating structure and its mechanical properties. That is why many studies have been devoted to modeling of the plasma jet and its fluctuations, the cold air engulfment[1-4], the particle injection[5], the interaction between particles and plasma[6], the flattening of droplets onto the substrate[7-13] and the coating formation[13]. In the most cases, the modeled particles were either metallic or ceramic. On the other hand, cermet composite coatings have interesting properties such as the improvement of the hardness of pure metals and the reduction of the brittleness of ceramic coatings. When both metal and ceramic particles are injected separately in the plasma jet, the behaviors of these two materials are quite different and simulations give information about their flattening and layering onto the substrate[14]. Unfortunately very few studies[15, 16] where devoted to cermet particles made by mechanofusion where a metal core is surrounded by a ceramic shell.

This paper presents results obtained when spraying, with a direct current Ar-$H_2$ plasma jet, stainless steel particles cladded through mechanofusion with an alumina shell. To better understand the experimental results related to the collection of particles in-flight, splats and coatings, models have been developed. The plasma jet temperature and velocity distributions are obtained with a 2D parabolic steady flow, a single particle trajectory is calculated by using 3D equations of Boussinesq-Oseen-Basset and the heat propagation phenomena within the two-layer particle has been taken into account including phase changes. The models allow understanding the convective phenomena induced within stainless steel particles by the plasma jet flow and the different steps of alumina and steel melting. If the force induced by the plasma at the particle surface has been calculated, the entrainment molten alumina (cermet particle) or the molten oxide (stainless steel particle) has, unfortunately, not modeled.

First the experimental conditions will be described. Second the models will be shortly presented. Third the results obtained successively with stainless steel particles then with stainless steel cladded by alumina will be described and discussed with the help of models.

## 2. Experiments:
### 2.1. Powders

The cladded stainless steel particles used in this study, are produced by mechanofusion. This process is used to manufacture composite powders starting from two or more raw powdered materials having different grain sizes. 316L stainless steel particles (Fe + 19 wt % Cr + 11 wt % Ni) with a size distribution from 45 to 100 µm, with a mean size of 58 µm, were cladded with pure α-alumina (99.99 %) particles of mean size 0.6 µm by the mechanofusion process. In this process, the raw particles are mixed intensively and subjected to a powerful compression, resulting in various strong mechanical forces generating thermal energy. Thereby, the material with the lowest melting temperature is heated to a plastic state and becomes the core of a composite particle to which the material with the highest melting temperature is cladded as a surrounding shell. The composite particles (Fig. 1) have a spherical shape and their mean diameter, after sieving, is ~ 65 µm (size between 50 and 80 µm). They are composed of a stainless steel core coated by an alumina shell of mean thickness 2 - 4 µm [17].

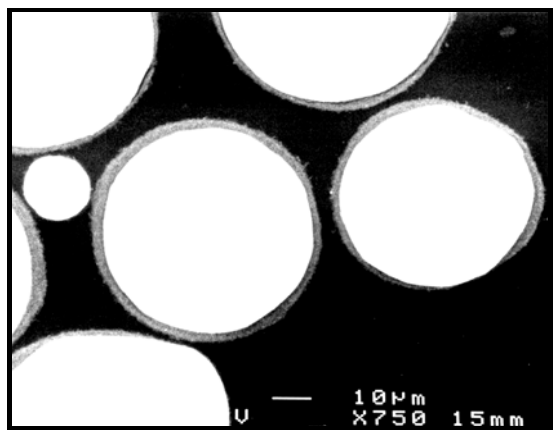

**Fig. 1 : Cross section of mechanofused stainless steel/$Al_2O_3$ powders**

Pure stainless steel particles (those forming the core of the mechanofused ones) were also sprayed in the same conditions for comparison. The properties presented in Table 1 were used for the calculations.

Table 1: Properties of stainless steel and alumina used in the calculations

|  |  | $\kappa$ [W·m$^{-1}$·K$^{-1}$] | $c_p$ [J·kg$^{-1}$·K$^{-1}$] | $\rho$ [kg·m$^{-3}$] | $T_m$ [K] | $T_b$ [K] | $L_f$ [J/kg] |
|---|---|---|---|---|---|---|---|
| **Stainless steel** | *Liquid* | 30 | 400 | 7 960 | 1 810 | 3135 | 2.472x10$^5$ |
|  | *Solid* | 40 | 1000 | 7 960 |  |  |  |
| **Alumina** | *Liquid* | 5 | 1200 | 3990 | 2 326 | 3800 | 1.093 x10$^6$ |
|  | *Solid* | 6.3 | 1888 | 3990 |  |  |  |

### 2.2. Spray conditions

Air plasma spraying (APS) was performed with a conventional d.c. plasma gun that has been described previously [18]. The cathode was 7 mm in diameter with a conical tip (cone angle of 40 °) and was made of thoriated tungsten (2 wt % ThO$_2$). The torch anode/nozzle was made of OFHP copper and had a conical shape (same cone angle as that of the cathode) followed by a cylindrical duct **7** mm in internal diameter (i.d.) and 28 mm in length.

The plasma jet parameters are shown in Table 2.

Table 2: Plasma jet parameters

| Arc current | [A] | 500 |
|---|---|---|
| Voltage | [V] | 57 |
| Torch thermal efficiency | [%] | 56 |
| Argon flow rate | [slm] | 53 |
| Hydrogen flow rate | [slm] | 7 |

Both powders were injected perpendicularly to the plasma jet axis through a 1.8 mm i.d. injector located 3 mm upstream of the torch nozzle exit. For both powders the Ar carrier gas flow rate was optimized at 5.5 slm in order that the mean particle trajectory made an angle of 3.5 ° with the torch axis.
The substrates were made of AISI 304L steel disks 5 mm thick and 25 mm in diameter. They were mounted on a cylindrical sample holder (external diameter : 110 mm) rotated at about 3 rps, while the plasma torch, the axis of which was at 90° to the sample holder axis, moved back and forth at 25 mm/s.
In collecting splat samples, only one pass of the torch was performed at the lowest powder feed rate (~ 20 - 25 g/h), to avoid overlapping of the splats.

For coatings, the feed rate of the injected powder was ~ 0.6 kg/h and the deposition time 10 minutes. The substrates were preheated to 200°C, by the plasma gun without powder feeding. Powder injection started when the required preheating temperature was reached. The preheating time was limited to ~ 200 s to minimize substrate oxidation [19].

During preheating and deposition, the surface temperature of the substrate or coating was continuously monitored by an infrared pyrometer ($\lambda$ = 6.16 µm, $\varepsilon$ = 0.85) which controlled the cooling air flow rate.

As the splat morphology strongly depends on the topography of the substrate surface [21], substrates used for splat collection were polished with SiC paper, then with diamond abrasives and finally ultrasonically cleaned in acetone. The resulting roughness (measured by Atomic Force Microscopy) was Ra ~ 0.05 µm.

Substrates used for coatings were grit blasted with white corundum with a mean diameter of 400 µm (grit blasting nozzle internal diameter of 8 mm, suction type machine, pressure of $4.10^5$ Pa, grit blasting distance ~ 150 mm and blasting surface rate of ~ 10 $cm^2$/s). After grit blasting, samples were blown with compressed air and then cleaned ultrasonically in acetone. The resulting mean roughness was Ra ~ 6.7 [± 0.3] µm.

### 2.3. Particles collection in-flight

To collect particles in flight, the plasma jet was oriented towards a nearly concentric cylinder 130 mm in diameter and 650 mm in length as described elsewhere [20]. The front of the cylinder was closed with a water cooled flange containing a 16 mm diameter inlet hole. The cylinder was positioned so that this hole was at a distance between 50 and 100 mm (the latter being the usual "stand off" distance) from the plasma gun exit and the hole was centered to the axis of the spray cone issuing from the plasma gun. The molten particles entering through this hole were quenched by argon jets (total flow rate of 50 slm) positioned just behind the collecting cylinder inlet. With this apparatus, more than 90 wt % of the particles composing the deposit were collected.

### 2.4. Characterization and analyses

Powders, splats and coatings were examined by optical microscopy (OM) and scanning electron microscopy (SEM). Since the contrast in backscattered electron (BE) images is related to the local atomic weight, a qualitative evaluation of phase mixing was performed by means of BE images. Energy dispersive spectroscopy (EDS) was used to analyze the distribution of the chemical elements. X-ray diffraction was used to determine the phases present within the starting and collected powders, as well as within coatings. The patterns were recorded on a Siemens diffractometer and compared with JCPDS files.

The profiles of splats were observed with a profilometer. Twenty profiles of the same splat were examined to determine the thickness distribution.

Coating microhardness measurements were performed on polished cross sections using a Vickers hardness (HV) tester with a load of 5 N applied over 15 seconds. Twenty measurements of microhardness were performed to determine mean values.

The oxide content within particles (nodules and cap) was measured by image analysis techniques (Paint Shop Pro - 7 and Matrox inspector - 2.1) giving the percentage of surface area of oxides as nodules and cap. The oxides were characterized by their black color, it was checked previously by EDS that those black area contained Fe, Cr, Ni, Si and O. Particles, mounted in epoxy, were polished to conduct scanning electron microscopy of their cross sections. The gray-shaded micrographs were transformed into dichromatic images by graphic software. For every batch 25 micrographs (about 70 particles) were treated to record surface area value of oxides nodules, oxide cap and metal.

## 3. Model:
### 3.1. Plasma particle interaction

The plasma jet is modeled with Jet&Poudres[21], a 2-D parabolic flow derived from the Genmix 2-D axisymmetric algorithm with the proper thermodynamic and transport properties of plasma gases computed with T&TWinner data bank[22]. The 3D equations of Boussinesq-Oseen-Basset allow determining the trajectory of a single particle with a given injection vector[23], its velocity as well as the plasma temperature and velocity "seen" by it, along its trajectory. Thus it allows calculating the Reynolds number imposed by the gas flow to the particle (see eq.1) as well as the cinematic viscosities of the plasma and the stainless steel particle assumed to have a uniform temperature (lumped model).

$$\text{Re} = \frac{\rho_p (v_\infty - v_p) d_p}{\mu_\infty} \qquad \text{eq. 1}$$

Compared to previous works, the heat transfer propagation phenomena within the two-layer particle, including phase changes have been taken into account as well as the existence of a constant or variable thermal contact resistance (TCR) between both materials [23]. However the interaction between the plasma flow and the liquid particle (convective movements induced by the gas flow inside the liquid particle or the entrainment of liquid oxides at the surface of the molten steel) is not modeled. For stainless steel particle, as its best number (see eq.2 and eq.3) is below 0.01 the heat propagation phenomenon within it has been neglected and a lumped heat transfer model used.

$$Bi = \frac{\overline{\kappa_\infty}}{\kappa_p} \qquad \text{eq. 2}$$

where $\kappa_p$ is the thermal conductivity of the particle and $\overline{\kappa_\infty}$ the mean integrated thermal gas conductivity [24].

$$\overline{\kappa_\infty} = \frac{1}{T_\infty - T_p} \int_{T_p}^{T_\infty} \kappa_\infty dT \qquad \text{eq. 3}$$

The numerical solution of the heat transfer[24-26] is based on the enthalpy model and the discretization equations are obtained by finite volumes in second order in time and space.

The variable thermal contact resistance (TCR) has been introduced in the model in order to follow the evolution of the physical contact between both layers. This contact varies with temperature due to the difference in thermal expansion coefficients between both materials, that of the ceramic shell being significantly lower than that of the metal core ($8 \cdot 10^{-6}$ K$^{-1}$ for alumina against $12 \cdot 10^{-6}$ K$^{-1}$ for steel). Upon heating this difference may lead to the solid ceramic shell cracking. Without any information about TCR in these mechanofused particles, the variable TCR has been assumed to vary linearly with the mass fraction of solid alumina material ($fs_m$)

$$(TCR)_{variable} = fs_m \, TCR_s + (1 - fs_m) \, TCR_l \qquad \text{eq. 4}$$

$f_{sm}$ is defined by the average mass fraction of alumina and steel near the interface.
. To check the influence of $TCR_s$ on the mechanofused particle heating, different values of $TCR_s$ have been taken from a poor contact: $10^{-5}$ m$^2 \cdot$K W$^{-1}$ to an almost perfect contact $10^{-8}$ m$^2 \cdot$K W$^{-1}$.

The values of thermal ($h_\infty$ [W·m$^{-2}$·K$^{-1}$]) and mass ($h_m$ [m·s$^{-1}$]) transfer coefficients are calculated according to the semi empirical correlations of Ranz and Marshall[26] established for liquid droplets in translation movement, and using the Nusselt (Nu) and the Sherwood (Sh) numbers

$$\text{Nu} = 2 + 0.6\, \text{Re}^{0.5}\, \text{Pr}^{0.33} \qquad \text{eq. 5}$$

$$\text{Sh} = 2 + 0.6\, \text{Re}^{0.5}\, \text{Sc}^{0.33} \qquad \text{eq. 6}$$

where Pr, Re and Sc are Prandtl, Reynolds and Schmidt dimensionless numbers respectively defined as:

$$\text{Pr} = \frac{\mu_\infty c_{p\infty}}{\kappa_\infty} \qquad \text{eq. 7}$$

$$Sc = \frac{\mu_\infty}{\rho_\infty D_\infty} \qquad \text{eq. 8}$$

with, $\kappa_\infty$ (W·m$^{-1}$·K$^{-1}$) the integrated average thermal conductivity of the plasma, $c_{p\infty}$ (J·kg$^{-1}$·K$^{-1}$) the specific heat of the particle at constant pressure, $D_\infty$ (m$^2$·s) the diffusion coefficient of the volatile species in the plasma, $\rho_\infty$ (kg·m$^{-3}$) and $\mu_\infty$ (Pa·s) the specific mass and viscosity of the plasma.

For the in-flight particles in the plasma, correction factors have been introduced to account for the temperature gradient within the boundary layer surrounding the particle, as well as the presence of the vapor cloud due to alumina evaporation[28, 29]. For example the thermal transfer coefficient was calculated as:

$$h_\infty = \frac{k_\infty \overline{\text{Nu}}}{d_p} = \frac{\overline{k_\infty}}{d_p} \left[ (2 + 0.6\, \text{Re}^{0.5}\, \text{Pr}^{0.33}) \left( \frac{(\rho\mu)_\infty}{(\rho\mu)_s} \right)^{0.6} \left( \frac{c_p(T_\infty)}{c_p - T_{ps}} \right)^{0.38} f_{vap} \right] \qquad \text{eq. 9}$$

with $\quad f_{vap} = \frac{\dot{m}_v}{d_p} \frac{c_{p\infty}}{\pi \overline{k_\infty}} \left[ \exp\left( \frac{\dot{m}_v}{d_p} \frac{c_{p\infty}}{\pi \overline{k_\infty}} \right) - 1 \right]^{-1} \qquad \text{eq. 10}$

where $\dot{m}_v$ (kg·s$^{-1}$) is the evaporated mass flow rate,

If within the stainless steel particle the temperature is constant along its radius, within the two-material particle exists two temperature gradients, i.e. non uniform temperature distributions enlarged by the phase changes. The phase transitions were considered as a Stefan problem without natural convection within the particle due to its size (< 100 µm) and its very short residence time (~ 1ms) in the plasma (compared to the convective movement characteristic time ~ 1s).

Particle vaporization starts below the boiling temperature and the mass flux lost by the particle, controlled by the vapor diffusion through the boundary layer surrounding the in-flight particle, is given by [23]:

$$\overset{\circ}{m}_v = \pi d_p^2 \left( \rho_v D_v \right)_f \ln(1+B)\, Sh \qquad \text{eq. 11}$$

where $\rho_v$ (kg·m$^{-3}$) is the specific mass of the gas resulting from alumina vaporization, $D_v$ (m$^2$·s$^{-1}$) its diffusion coefficient, both calculated at the film temperature for Al2O3, which is the alumina vapor, diffusing in argon, at last B is the mass transfer number of Spalding defined by

$$B = \frac{Y_s - Y_\infty}{1 - Y_s} \qquad \text{eq. 12}$$

where $Y_s$ and $Y_\infty$ are the mass fraction of species evaporating in the boundary layer and those existing in the plasma gas, here supposed to be close to zero. $Y_s$ is given by

$$Y_s = \frac{M_v}{M_v + M_\infty \left( \frac{p_\infty}{p_v(T)} - 1 \right)} \qquad \text{eq. 13}$$

where $M_v$ (kg·mol$^{-1}$) is the molar mass of volatile species, $M_\infty$ that of the plasma and $p_\infty$ is the plasma pressure while $p_v$ is the saturated vapor pressure at the particle surface calculated according to Clausius-Clapeyron law assuming the gas is perfect[31].

Particles are considered to be spherical, and it is assumed that, for the cladded particles, both materials keep their specific masses during the solid-liquid phase transition. Their physical properties are temperature dependent and the mass transfer is supposed to be radial.
The boundary layer is supposed to be in thermodynamic equilibrium and without chemical reactions.
The particle trajectory is obtained by solving the momentum equation (eq.16) using a fourth-order Runge Kutta method:

$$\frac{d\vec{v_p}}{dt} = \frac{3}{4}\left(\frac{\rho_\infty}{\rho_p d_p}\right) C_D \left|\vec{v_\infty} - \vec{v_p}\right| \ast (\vec{v_\infty} - \vec{v_p}) + \vec{g} \qquad \text{eq. 14}$$

in this equation $C_D$ represents the drag coefficient, and v (m·s$^{-1}$) the velocity of the plasma (lower index $\infty$) and that of the particle (lower index p), g (m·s$^{-2}$) is the gravity acceleration negligible with d.c. plasma.
Of course, as for the heat transfer, $C_D$ is corrected to account for the steep temperature gradient within the boundary layer surrounding the particle as well as the Knudsen effect.

### 3.2. Stress induced by steel melting before alumina
The stress, when the iron core melts before alumina, can be calculated by using the expression of Lamé given by Timoskunko and Grodia [25] for a thin shell (e $\leq R_0/4$ where e is the shell thickness and $R_0$ the particle radius).

$$\sigma_m = \frac{\Delta p}{p_\infty}\left(\frac{R_0}{e}\right) \qquad \text{eq. 15}$$

where $\Delta p$ is the over pressure due to the volumetric expansion change when the phase change occurs

$$\frac{\Delta p}{p_\infty} = (\alpha_{SS} - 3 \lambda_{Al_2O_3} \cdot \Delta T) - 1 \qquad \text{eq. 16}$$

where $\alpha$ is the volumetric expansion coefficient of steel ($\alpha_{SS} = \frac{\rho_s - \rho_\ell}{\rho_\ell}$, $\rho$ being the specific mass while s and $\ell$ indexes stand for solid and liquid).

# 4. Experimental results and discussion according to calculation
## 4.1. Stainless steel particles
### 4.1.1 Collected along the optimum trajectory

Stainless steel particles, sprayed according to the conditions summarized in Table 2 were collected in-flight respectively at 50, 60, 80 and 100 mm downstream of the nozzle exit, the powder collector being centered on the mean particle trajectory.

At 50 mm most particles collected are similar to that presented in Fig. 2a the surface structure showing evidence (Fig. 2b) of possible convective movements and the presence of an oxide cap.

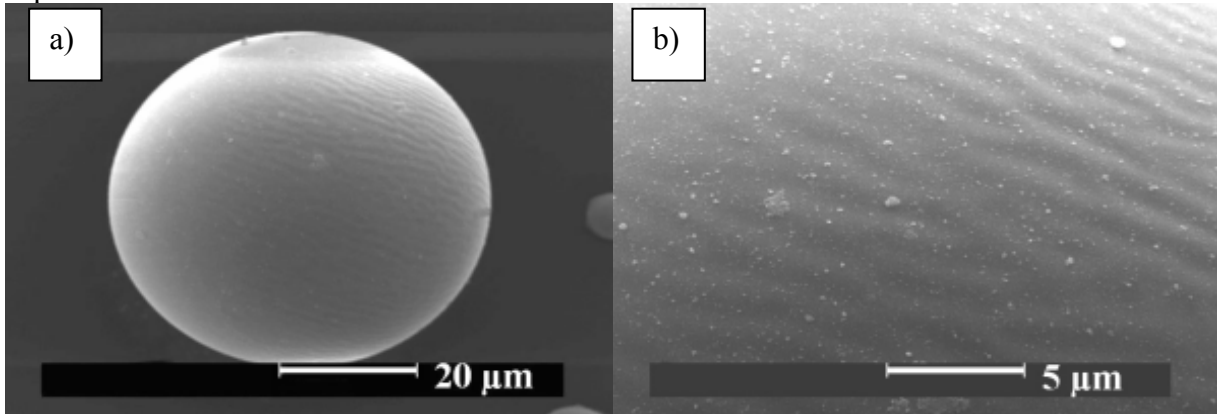

**Fig. 2 : Surface of a stainless steel particle collected in-flight: convective movement effects (stand off distance: 50 mm, I=500A, Ar-$H_2$=53-7 NL.min$^{-1}$, anode i.d.=7mm)[35] a) whole particle, b) enlargement**

The cross section of this particle (Fig. 3) shows inside it the existence of dark oxide nodules and outside an oxide cap. The EDS analysis of the white iron section has shown only Fe, Cr, Ni and Si, while the dark nodules exhibited the same elements plus oxygen. These nodules can be explained only if a convective movement inside the molten steel is induced by the plasma flow. For that it is necessary that the Reynolds number, Re, relative to the particle (see eq.1) is higher than 20 and that $v_g/v_p > 55$ where $v_g$ is the cinematic viscosity of the plasma ($v_g = \mu_g/\rho_g$ and $v_p = \mu_p/\rho_p$ where $\mu_g$ and $\mu_p$ are respectively the molecular viscosity of the plasma and molten particle while $\rho_g$ and $\rho_p$ are respectively their specific masses [31, 32].

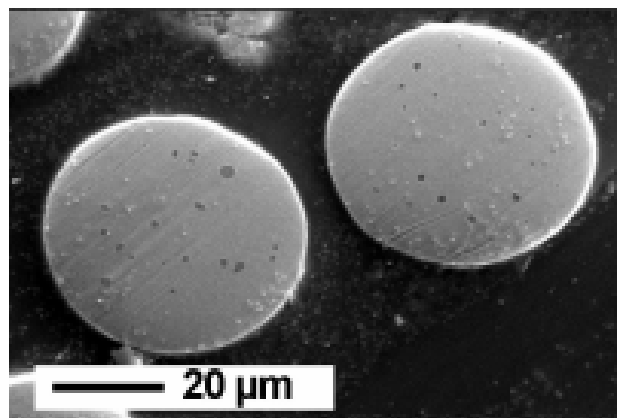

**Fig. 3 : Oxide nodules in cross section of a stainless steel particle collected in-flight (stand off distance: 50 mm, I=500A, Ar-$H_2$=53-7 NL.min$^{-1}$, anode i.d.=7mm)[35]**

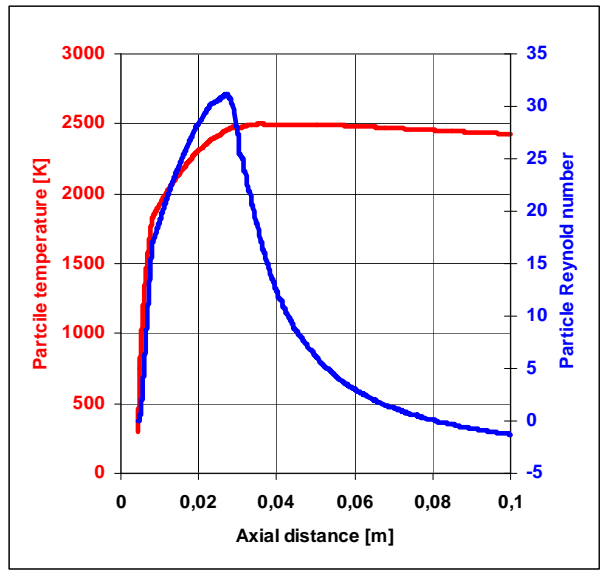

Fig. 4 represents the evolution of $Re_p$ along the optimum trajectory of the particle. Between about 12 and 35 mm downstream of the torch exit, the condition $Re_p > 20$ is for a 60 μm particle while $v_g/v_p > 55$ is satisfied as soon as the particle is melted i.e. 8 mm after its injection within the plasma jet (see Fig. 5). In such conditions Hill's vortex, induced within the molten metal, sweeps the oxide formed at the particle surface or the oxygen dissolved into the particle and renews continuously fresh metal at the surface.

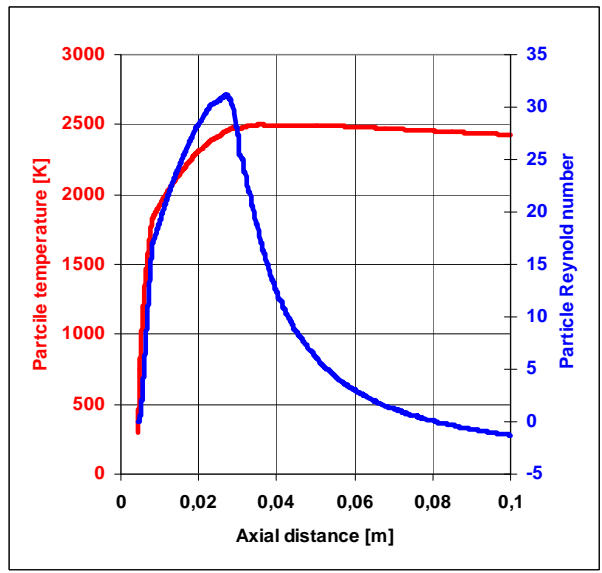

**Fig. 4 : Evolution of the Reynold number relative to the particle as well as the particle temperature along the 60 μm stainless steel particle trajectory in the Ar-H$_2$ (53-7 vol %) plasma (I = 500 A)**

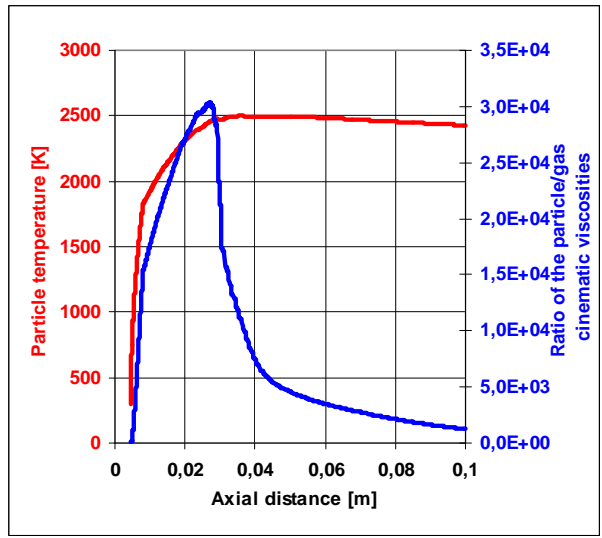

**Fig. 5 : Evolution of the ratio of the cinematic viscosities of the gas and particle as well as the particle temperature along its trajectory**

Fresh metal is always available on the particle surface resulting in an enhanced particle reactivity. Surface tension difference between liquid oxide and liquid metal formed promotes isolated spherical nodules of oxides. This result is similar to that obtained by Espié et al.[30] with low carbon steel particles. The oxide nodule percentage (in surface percentage) is $0.9 \pm 0.05$ % and all along the particle trajectory between 50 mm and 100 mm this value is constant. It confirms that the conditions for induced convection within the molten particles are no more fulfilled, as shown by Fig. 4 and thus nodule growing is impossible. Of course it is not the case with the oxide cap which at 50 mm represents about 0.9 % of the oxide surface, 2% at 70 mm and almost the same at 100 mm. However at that distance the total oxide content reaches 5 %. It has also to be noted that both at 50 mm and 100 mm stand off distances particles exhibit dark nodules, dark cap and broken oxide shell (see Fig. 6 obtained at 100 mm). Once, the convective movement within particles ceased, which occurs for distances higher than 35 mm for the 60 µm particles, conventional diffusion oxidation occurs. It is of course possible that surface oxidation values at the particle surface are slightly over estimated because oxidation may keeps going inside particles collector, because the argon blowing within the collector is not sufficient to prevent oxidation by diffusion when particles cool down at its bottom.

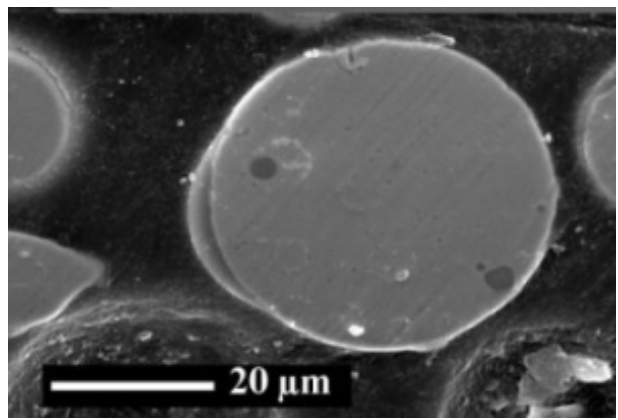

**Fig. 6 : Stainless steel particle collected in-flight**
**(stand off distance: 100 mm, I=550A, Ar-$H_2$=45-15 NL.min$^{-1}$, anode i.d.=6mm)**[35]

The interesting point is the oxide cap formation observed for the particles collected at different distances from the nozzle exit. After about 8 mm trajectory the 60 μm particle is fully melted but according to Fig. 4 convective phenomenon starts only at 12-13 mm. Thus an oxide layer is formed at the particle surface by diffusion. This oxide layer formation, close to the beginning of the trajectory, is highly probable because measurements performed by emission spectroscopy [33] have shown that in similar spray conditions at 10 mm, on the plasma jet axis the ratio O/Ar is 0.7 % and reaches 5 % at 30 mm. This is due to the engulfment process induced by the plasma flow exiting at high velocity (2000 m/s at the torch axis) in air, process promoted by arc root fluctuations and the induced piston flow. The plasma momentum density along the particle trajectory varies as shown in Fig. 7 and between 5 and 50 mm is higher than 5000 Pa, (up to about 40 000 Pa) and still over 1000 up to 70 mm Fig. 5. Such values are quite sufficient to entrain the light liquid oxide layer formed towards the leading edge of the particle, thus creating the oxide cap. After wards the oxide formed at the surface is no more entrained to the leading edge and stays around the particle.

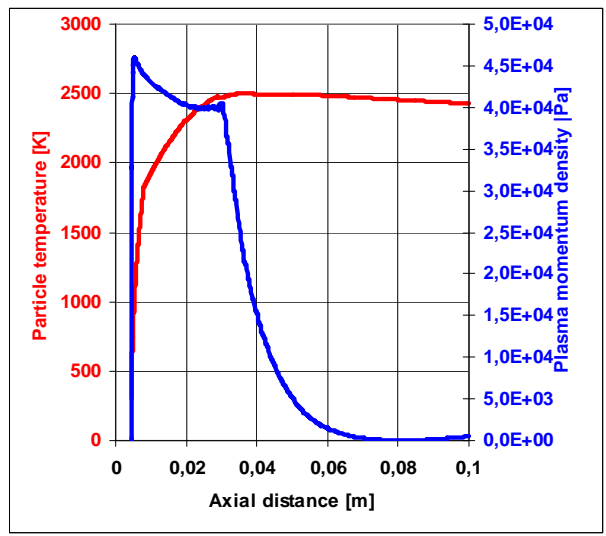

**Fig. 7 : Plasma momentum density along the particle trajectory**

### *4.1.2 Collected along a trajectory crossing the plasma jet*

When the 60 μm particle trajectory makes an angle of 10° with the torch axis the conditions for convection movements are not at all fulfilled and the momentum density is much lower. Thus it is not surprising that when collecting particles with the collector centered on this trajectory most particles exhibit neither cap nor nodules but an oxide shell around the particle (oxidation controlled by diffusion).

### **4.2. Stainless steel particles with an alumina shell**
#### *4.2.1 Experiments*

Experiments, when spraying with the same conditions, particles made of stainless steel cladded by an alumina shell, show that particles collected in flight are stainless steel spheres either with alumina caps (most of them) or alumina pieces distributed unevenly at their surface especially when collecting them along trajectories making angles > 10° relatively to the torch axis. In the first case, due to the force induced by the plasma flow at the surface of the particle (see Fig. 7), the molten alumina (lighter: 3 600 kg/m² than the stainless steel core:

7 900 kg/m²) is entrained at the leading edge of the quasi-spherical molten metal, resulting, after solidification, in the shape shown in Fig. 8. This is the same phenomenon than that observed with the oxide cap for stainless steel particles (collected at 50 mm, Fig. 2a and at 100 mm Fig. 6). It is worth noting that there is no difference for particles collected at 50 mm. Upon impact on the substrate these particles result in a stainless steel splat below the alumina one (see Fig. 9). Particles with unevenly distributed alumina pieces at their surface (see Fig. 8) result in splats (see Fig. 9, right upper part) essentially made of stainless steel with some pieces of solid alumina at their surface. These particles are probably resulting from poorly heat treated particles where the alumina shell has been broken into pieces by the expansion of the steel core either not yet melted or with a temperature slightly over the melting temperature (no entrainment of the molten alumina by the flow) and then solidified with their alumina shell partially broken.

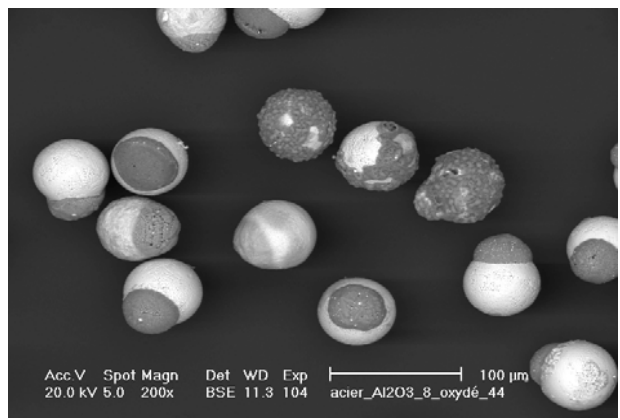

**Fig. 8: Plasma sprayed mechanofused stainless steel/alumina particles collected in flight in the Ar-H$_2$ (53-7 slm) plasma jet (I = 500 A, nozzle i.d. 7 mm, stand off distance of 100 mm): white color stainless steel, grey color alumina**

The question which is raised for particles with the cap is the following: is the alumina shell fully melted when the momentum density of the plasma along trajectory is very high (between 8 and 50 mm for values over 5 000)

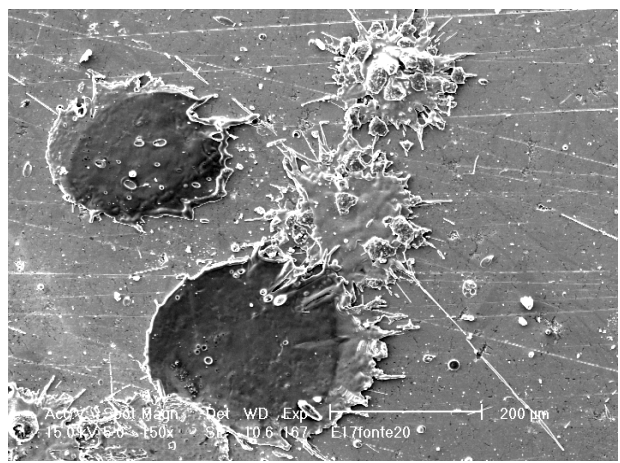

**Fig. 9: Splats of mechanofused stainless steel/alumina particles collected on a smooth cast iron substrate preheated at 200°C and plasma sprayed with the Ar-H$_2$, defined in** Erreur ! Source du renvoi introuvable.
**caption
(white color stainless steel, grey color alumina)**

*4.2.2 Heat and mass transfer modeling*

Comparatively to the mono-material particle, the two-layer one requires more parameters to be characterized such as ratios of layer densities and specific heats, the ratio of iron core radius to that of particle and at last the thermal contact resistance TCR between the alumina shell and the stainless steel iron core. The value of this TCR plays a drastic role on the heat transfer from the alumina to the iron. Unfortunately it is not possible to measure it and calculations can only give trends for different TCR values from a poor contact TCR = $10^{-5}$ $m^2 \cdot K \cdot W^{-1}$ to an almost perfect one close to $10^{-8}$ $m^2 \cdot K \cdot W^{-1}$. In addition the problem is made very complex by the drastic difference between the expansion coefficients of both materials.

The particle considered is 60 μm in diameter with an alumina shell 2 μm thick. It is injected to follow either an optimal trajectory obtained by adjusting its initial velocity to achieve a trajectory making an angle of 3.5° with the torch axis or a peripheral trajectory with an angle of 30° with the torch axis [36, 37].

The problem became physically complex and any method of numerical modeling and simulation have to follow this increasing nonlinearity. Particularly, the contact introduces a gap in ethe thermal field and phase change introduces a gap in the heat flux.

For solving the heat transfer coupled to phase change under the plasma environment, an enthalpy method successfully used for directional solidification [réf enthalpy] has been adapted to then spherical configuration. Specific development for the imperfect contact (alumina/steel) has been introduced.[Laraqui, Bouneder, ..]

*a/ - Particle heating along its optimum trajectory with constant TCR*

*i/- Heat transfer*

The calculation is first performed with a medium value and constant TCR of $10^{-6}$ $m^2 \cdot K \cdot W^{-1}$ and results are presented in Fig. 10 as function of the time spent by the particle along its trajectory (the corresponding positions along the particles trajectory are indicated below the time axis).

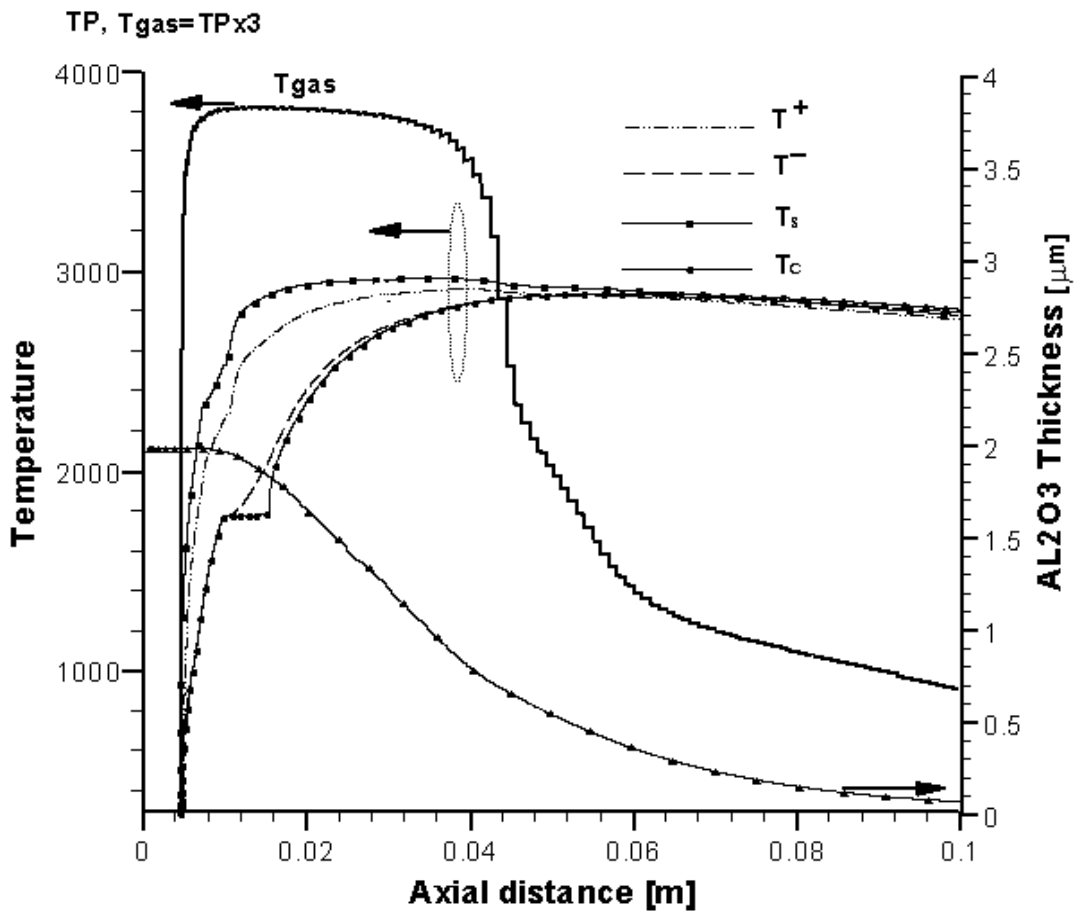

**Fig. 10:** Thermal history of stainless steel/alumina mechanofused particle ($d_p$ = 60 μm, alumina shell thickness 2 μm) along its optimal trajectory with a constant TCR = $10^{-6}$ m$^2$·K·W$^{-1}$ taking into account evaporation over melting point

In this graph are shown:
- the plasma temperature seen by the particle along its trajectory (scale of particle temperature divided by a factor 3),
- the surface and bottom temperatures of the alumina shell as well as the surface and center temperature of the stainless steel core,
- the progress of the alumina shell thickness due to its evaporation.

It can be noted that most important phenomena affecting the particle held at the 20 mm from the nozzle . the target situate at practically 100 mm. Thee alumina surface reaches its melting temperature first, while its bottom starts to melt about 20 μs later almost when steel starts melting (⊔ 10mm from the nozzle). It is also worth noting that steel core center starts melting almost at the same time as its surface but the latent heat dissipation takes almost 400 μs. Evaporation becomes noticeable as soon as the alumina surface temperature reaches 2 700 K and keeps going close to the substrate where alumina and steel impact almost at the same temperature around2 800 K with both materials fully melted (of course in this calculation the molten alumina entrainment by the flow is not considered). Alumina is fully melted after 11 mm trajectory which explains why the cap can be formed.

It is interesting to make the same calculation as that presented previously for a particle in fight but assuming that the particle can be heated up to its boiling point without evaporation.

Results are presented in Fig. 11. It can be seen that the alumina shell surface temperature reaches its boiling point (~ 3 800 K) in the plasma core almost when the particle trajectory reaches 30mm in. When taking account evaporation (see Fig. 10) the maximum surface temperature is about 2950 K. Thus evaporation starting below vaporization temperature limits the temperature increase through the evaporated mass as indicated in literature Delluc[24]. But here again alumina is fully melted after 11 mm trajectory.

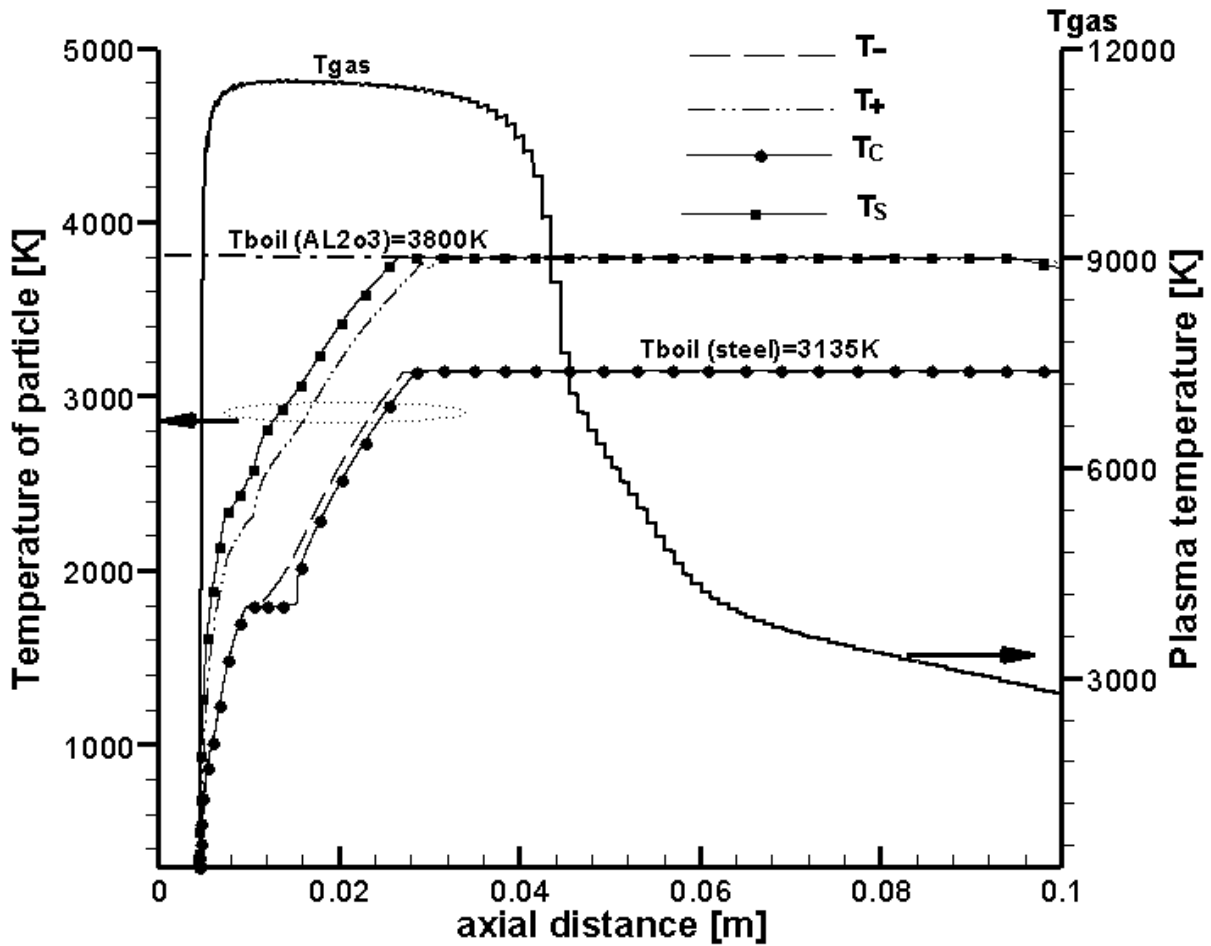

**Fig. 11: Thermal history of stainless steel/alumina mechanofused particle ($d_p$ = 60 μm, alumina shell thickness 2 μm) along its optimal trajectory with a constant TCR = $10^{-6}$ m²·K·W⁻¹ but with no evaporation before reaching alumina boiling temperature.**

With larger TCR = $5.10^{-6}$ m²·K·W⁻¹ the particle behavior is quite different (see Fig. 12). The temperature jump at the interface alumina/SS reaches 2 300 K and almost all the plasma energy is trapped within alumina accelerating its melting and its evaporation. At x = 45 mm from the nozzle almost all the alumina is evaporated while the iron, protected by the alumina evaporation, is not completely molten and its temperature at the contact practically has only reached about 1 870 K. Of course the calculation was stopped at that time. When using a larger TCR ($10^{-6}$ instead of $5x10^{-6}$ m²·K·W⁻¹ the alumina shell is not totally evaporated and the iron melting starts at distances over 15 mm. In this case, the entrainment of the alumina shell is probably not possible (friction of liquid alumina on solid iron is probably higher than that of liquid steel) and it could explain the existence of particles covered with a broken alumina layer (see Fig. 8).

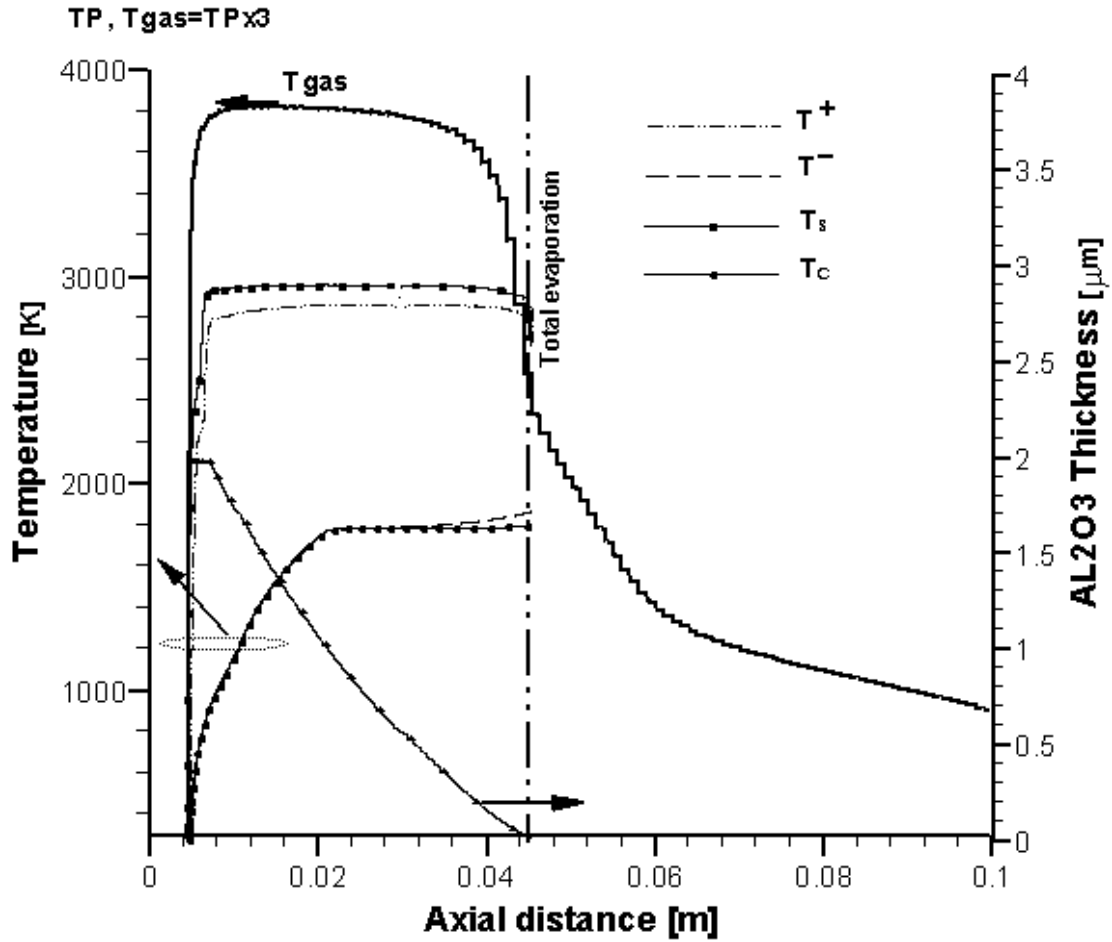

**Fig. 12:** Thermal history of stainless steel/alumina mechanofused particle ($d_p = 60$ μm, alumina shell thickness 2 μm) along its optimal trajectory with a constant TCR = $10^{-5}$ m$^2$·K·W$^{-1}$ taking into account evaporation over the melting point.

When considering a low TCR = $10^{-7}$ m$^2$·K·W$^{-1}$ results are similar to those of Fig. 10, except that steel melting temperatures is reached earlier (almost no difference for alumina) and the final thickness of the alumina shell at impact is larger than that corresponding to TCR=$10^{-6}$ (about 0.4 μm).

*ii/- Expansion mismatch problem*

- First it is interesting to see what will happen when both materials remain solid. The stress imposed to the alumina shell is written:

$$\sigma_{A\ell_2O_3} = E_{A\ell_2O_3} \left( \left(\frac{\Delta r}{r}\right)_{SS} - \left(\frac{\Delta r}{r}\right)_{A\ell_2O_3} \right) \quad \text{eq. 17}$$

The Young's modulus of alumina has been considered to be only one tenth of that of bulk alumina (372 GPa) because it consists only of fine alumina particles ($\bar{d} = 0.6$ μm) agglomerated by mechanofusion. It has been assumed that the mechanofused particle follows the optimum trajectory with TCR = $10^{-6}$ m$^2$·K·W$^{-1}$. At t = 0.25ms where the steel core surface temperature is about 1 500 K, while the alumina surface reaches its melting temperature, according to eq.(8) $\sigma_{Al_2O_3}$ = 37. 2 MPa (below its maximum tensile stress of about 80 MPa). Thus alumina will not crack. Of course if $E_{Al_2O_3}$ is assumed to be 1/3 of that of bulk material $\sigma_{Al_2O_3}$ = 124 MPa, cracks appear !

In at case, for a 2μm thick alumina shell with TCR = $10^{-7}$ m$^2$·K·W$^{-1}$ stainless steel melts after about 0.25 ms and eq.16 results in $\sigma_m$ = 137 MPa (with $E_{Al_2O_3}$ = 37.2 GPa). Thus the alumina shell will be broken. However this phenomenon will occur a few tens μs before shell melting and it will occur only if TCR ≤ $10^{-7}$ m$^2$·K·W$^{-1}$.

### b/ - Particle heating along its optimum trajectory. Variable TCR

Previous results with a constant TCR have shown that the alumina shell melting always occurs before that of stainless steel. As the contact Al$_2$O$_3$/SS is increased by this melting, a variable TCR has been used, its value being assumed to vary linearly with the mass fraction of solid near the contact interface (see eq.4). It has been assumed in Fig. 13 that for solid alumina TCR$_s$ = $10^{-6}$ m$^2$·K·W$^{-1}$ while for liquid one TCR$_l$ = $10^{-8}$ m$^2$·K·W$^{-1}$.

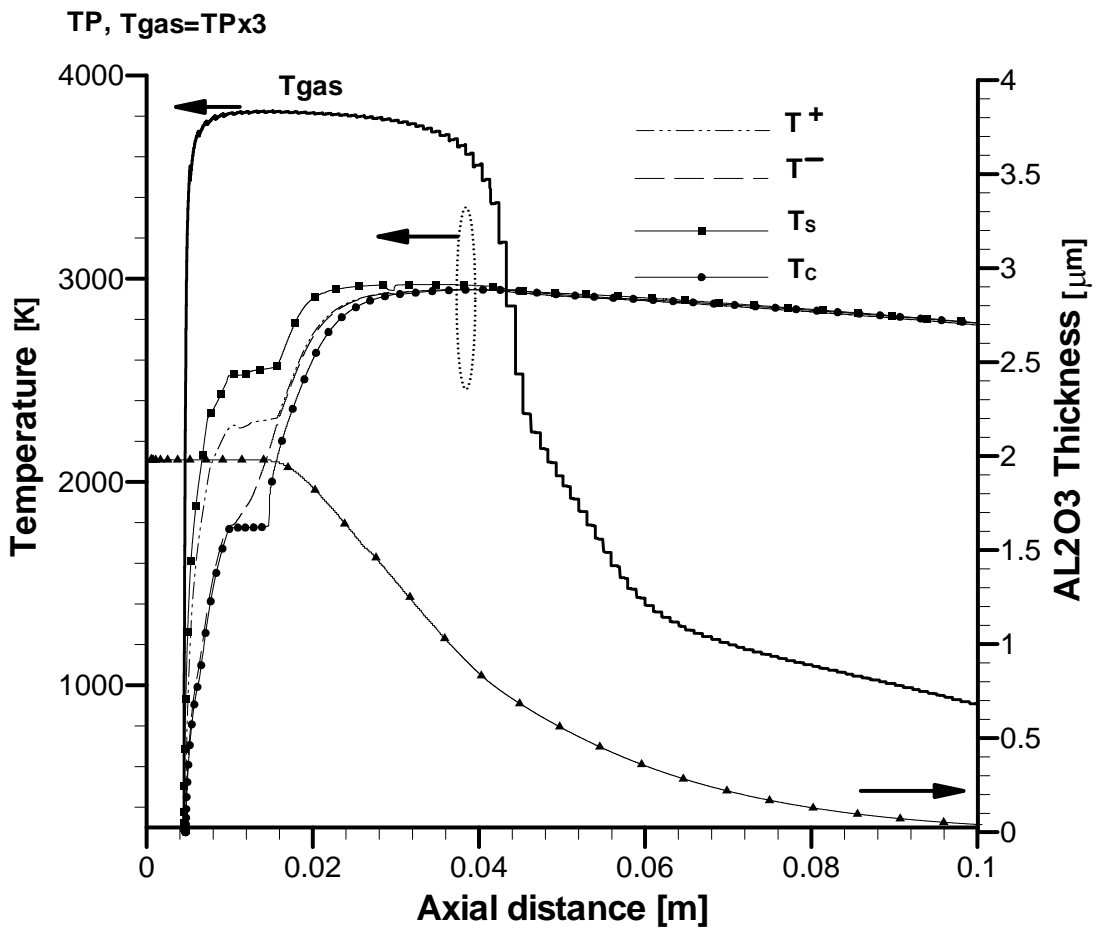

**Fig. 13: Thermal history of stainless steel/alumina mechanofused particle (d$_p$ = 60 μm, alumina shell thickness 2 μm) along its optimal trajectory with a variable TCR from 1.10$^{-6}$ to 10$^{-8}$ m$^2$·K·W$^{-1}$ taking into account evaporation before reaching the melting temperature.**

Compared to Fig. 10, where TCR = $10^{-6}$ m$^2$·K·W$^{-1}$, the temperature jump is reduced as soon as the whole alumina particle is liquid. As with the constant TCR the alumina shell is melted before the stainless steel but its evaporation starts later (39 mm against 34 mm).

Table 3 summarizes the differences between constant and variable TCRs with different starting values of TCR. These differences are characterized by the maximum temperature jump as well as the melting times respectively of alumina, $\tau_{Al_2O_3}$ and iron, $\tau_{Fe}$. This table illustrates the drastic importance of TCR. With the decrease of contact resistance TCR, $\Delta T_{max}$ decreases as well as the melting time of iron, $\tau_{Fe}$, while that of alumina, $\tau_{Al_2O_3}$, increases. For a variable TCR, starting from the same value as the constant one and finishing at $10^{-8}$ m$^2$·K·W$^{-1}$ when the alumina shell and stainless steel near the contact interface is fully melted, $\Delta T$ is lower than with constant TCR, the melting time of iron is shorter while that of alumina is unchanged as it could be expected.

**Table 3: Influence of the thermal contact resistance on characteristics temperature jumps and melting times of 60 μm alumina/steel particle (alumina shell thickness 2 μm) along its optimum trajectory.**

| TCR (m$^2$·K·W$^{-1}$) | | $10^{-5}$ | $5 \cdot 10^{-6}$ | $2.5 \cdot 10^{-6}$ | $10^{-6}$ | $10^{-7}$ | $10^{-8}$ |
|---|---|---|---|---|---|---|---|
| **Case (A)** TCR = C$^{ste}$ | $\Delta T_{max}$ (K) | 2 272 | 1 943 | 1 454 | 728 | 97 | 15 |
| | $\tau_{Fe}$ (ms) | - | 0.436 | 0.339 | 0.302 | 0.3107 | 0.288 |
| | $\tau_{Al2O3}$ (ms) | 0.200 | 0.204 | 0.220 | 0.256 | 0.288 | 0.328 |
| **Case (B)** TCR var. from different values to $10^{-8}$ (m$^2$·K·W$^{-1}$) | $\Delta T_{max}$ (K) | 2085 | 1621 | 1257 | 660 | 86 | 15 |
| | $\tau_{Fe}$ (ms) | 0.448 | 0.343 | 0.311 | 0.304 | 0.288 | 0.288 |
| | $\tau_{Al2O3}$ (ms) | 0.200 | 0.204 | 0.220 | 0.255 | 0.324 | 0.328 |

However it must be underlined that in all cases considered alumina shell melting occurs during the first millimeters along the particle trajectory (between 8 and 15 mm). It means that the molten alumina shell is submitted to a high momentum density of the plasma flow entraining it to the leading edge of the in-flight particle.

### c – Particle heating along its peripheral trajectory. Variable TCR

Fig. 14 represents the particle behavior for a TCR starting at $10^{-5}$ m$^2$·K·W$^{-1}$ and finishing at $10^{-7}$ m$^2$·K·W$^{-1}$. As it could be expected the particle behavior is quite different from that along the optimal trajectory: first its residence in the plasma core is shorter and when exiting the plasma core the alumina shell solidifies fully whilethe metallic core remains in the transition state solid/liquid at its solidification temperature, up to the impact.

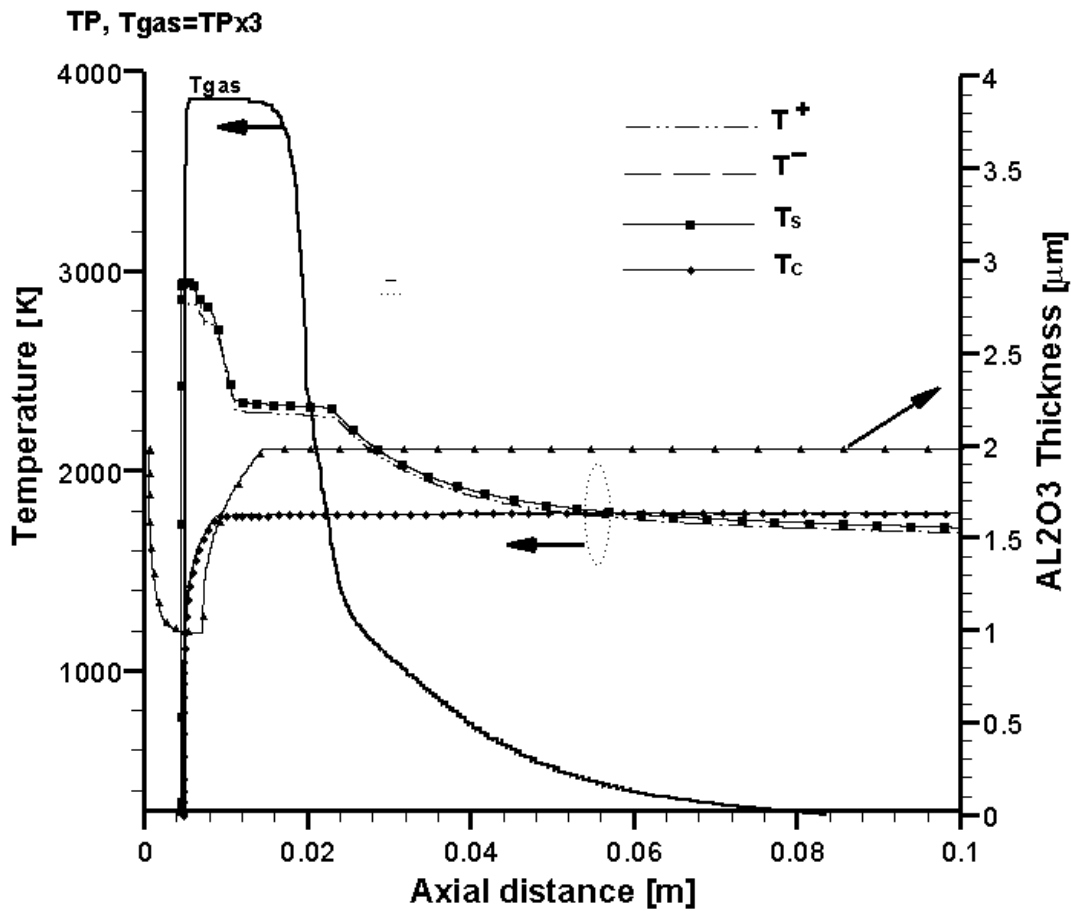

**Fig. 14:** Thermal history of stainless steel/alumina mechanofused particle ($d_p = 60$ μm, alumina shell thickness 2 μm) along its peripheral trajectory with a variable TCR from $1.10^{-5}$ to $10^{-7}$ m$^2\cdot$K$\cdot$W$^{-1}$ taking into account evaporation before reaching the melting temperature.

The resolidified alumina shell is probably broken. However when taking s TCR variable between TCRs=$10^{-6}$ and TCRl=$10^{-8}$ m$^2\cdot$K$\cdot$W$^{-1}$, both materials are melted during their travel in the plasma core and afterwards cool down slowly, reaching the substrate fully melted slightly over 2400 K. Probably in this case due to the much lower plasma velocity in the core fringes and the plasma plume, the molten alumina shell may not be entrained by the flow as for particles traveling in the plasma core with an optimum trajectory.

**Conclusion**:

This study is devoted to the modeling of the thermal transfer between a d.c. plasma jet and a single particle of alumina/steel elaborated by mechanofusion. Results are compared to the out look of particles collected in-flight when air plasma sprayed. The computations are related to an iron particle core (58 μm in diameter) surrounded by a 2 μm thick alumina shell in an argon-hydrogen plasma (Ar/H$_2$: 53/7 slm, 500 A, torch nozzle i.d. 7 mm, effective power 28.5 kW) while for experiments particle sizes were between 50 to 63 μm. Experimentally two types of particles were collected in flight at the stand-off distance of 100 mm: stainless steel particles with either an alumina cap (most of them) or with alumina pieces unevenly distributed at their surface. The alumina cap resulted from the entertainment by the plasma flow of the fully melted alumina shell surrounding the molten steel core.

Calculations have shown the drastic influence of the thermal contact resistance TCR between the alumina shell and the steel core. Unfortunately it could not be measured and thus different values of TCR have been tested.

Whatever may be the value of TCR the alumina shell melts before the steel core when particles follow an optimum trajectory making an angle of 3.5° with the torch axis. When TCR is high ($10^{-5}$ m$^2$·K·W$^{-1}$) the composite particle of 60 μm in diameter and 2 μm of alumina shell melt and evaporate very quickly and probably the particle does not exhibit a long life in the plasma. Evaporation of the alumina shell limits its maximum temperature to about 3 000 K. To account for the alumina shell melting the TCR has been supposed to vary and become lower when the alumina is liquid. This assumption makes the melting time of steel shorter than that obtained for a constant TCR and reduces the temperature jump between alumina and steel. The expansion mismatch between alumina and steel has also been taken into account. With a low Young's modulus of alumina (one tenth of that of bulk material because the alumina particles of the shell are only agglomerated by mechanofusion) the stress imposed to the alumina shell is not sufficient to break it before its melting. On the contrary if steel would melt before alumina, the alumina shell would be broken, which never occurs upon heating.

If the particle trajectory is not optimum, once the particle is no more within the jet core alumina may freeze before the particle reaches the substrate while the steel core is still in a molten state. In that case, if the alumina shell has not been entrained at the surface of the steel core by the flow, it will be broken into pieces before impact. It might be the case because the alumina temperature is slightly over its melting one thus resulting in a high viscosity of the liquid alumina which, with the much lower flow velocity along non-optimum trajectories, probably impedes its entrainment at the melted steel core surface.